\documentclass[10pt,british,oneside,onecolumn,a4paper]{article}
\usepackage{graphicx}
\usepackage[utf8]{inputenc}
\usepackage{mathptmx}
\usepackage{tabularx}
\usepackage{ragged2e}
\usepackage[singlelinecheck=false]{caption}
\usepackage[T1]{fontenc}
\usepackage[numbers]{natbib}
\usepackage{amsmath}

\usepackage{array}
\usepackage{rotating}
\usepackage{wrapfig}

\usepackage{todonotes}

\RequirePackage[blocks]{authblk}
\usepackage[british]{babel}
\makeatletter
\let\ps@plain\ps@empty
\def\@xivpt{14bp}

\def\@sect#1#2#3#4#5#6[#7]#8{%
  \ifnum #2>\c@secnumdepth
    \let\@svsec\@empty
  \else
    \refstepcounter{#1}%
    \protected@edef\@svsec{%
      \ifnum #2<4
        \hb@xt@10mm{\csname the#1\endcsname}\relax
      \else
        \hb@xt@12mm{\csname the#1\endcsname}\relax
      \fi}%
  \fi
  \@tempskipa #5\relax
  \ifdim \@tempskipa>\z@
    \begingroup
      #6{%
        \@hangfrom{\hskip #3\relax\@svsec}%
          \interlinepenalty \@M #8\@@par}%
    \endgroup
    \csname #1mark\endcsname{#7}%
    \addcontentsline{toc}{#1}{%
      \ifnum #2>\c@secnumdepth \else
        \protect\numberline{\csname the#1\endcsname}%
      \fi
      #7}%
  \else
    \def\@svsechd{%
      #6{\hskip #3\relax
      \@svsec #8}%
      \csname #1mark\endcsname{#7}%
      \addcontentsline{toc}{#1}{%
        \ifnum #2>\c@secnumdepth \else
          \protect\numberline{\csname the#1\endcsname}%
        \fi
        #7}}%
  \fi
  \@xsect{#5}}
\renewcommand\LARGE{\@setfontsize\LARGE{16}{20}}
\def\abstract#1{\def\@abstract{#1}}
\def\abstractEn#1{\def\@abstractEn{#1}}
\def\titleEn#1{\def\@titleEn{#1}}
\def\@maketitle{%
  \newpage
  \null
  \let \footnote \thanks
    {\LARGE\bfseries\RaggedRight \@titleEn \par}%
    \vskip 1\baselineskip%
    {\normalsize
      \@author\par}%
    \vskip 2\baselineskip%
    {\section*{Abstract}
      \@abstractEn}%
  \par
  \vskip 3\baselineskip}

\renewcommand\section{\@startsection {section}{1}{\z@}%
                                   {-3.5ex \@plus -1ex \@minus -.2ex}%
                                   {\baselineskip}%
                                   {\normalfont\Large\bfseries\RaggedRight}}
\renewcommand\subsection{\@startsection{subsection}{2}{\z@}%
                                     {\baselineskip}%
                                     {1ex}%
                                     {\normalfont\large\bfseries\RaggedRight}}
\renewcommand\subsubsection{\@startsection{subsubsection}{3}{\z@}%
                                     {1\baselineskip}%
                                     {3bp}%
                                     {\normalfont\normalsize\bfseries\RaggedRight}}
\renewcommand\paragraph{\@startsection{paragraph}{4}{\z@}%
                                    {1\baselineskip\@plus1ex \@minus.2ex}%
                                    {3bp}%
                                    {\normalfont\normalsize\RaggedRight}}
\renewcommand\subparagraph{\@startsection{subparagraph}{5}{\parindent}%
                                       {3.25ex \@plus1ex \@minus .2ex}%
                                       {-1em}%
                                      {\normalfont\normalsize\bfseries\RaggedRight}}
\parindent\p@
\makeatother
\DeclareCaptionLabelSeparator{enskip}{\enskip}
\titleEn{Design of a 5G Ready and Reliable Architecture for the Smart Factory of the Future}

\author{Mathias Strufe\textsuperscript{*}, Michael Gundall\textsuperscript{*}, Hans D. Schotten\textsuperscript{*\dag}, Christian Markwart\textsuperscript{\ddag}, Rakash S. Ganesan\textsuperscript{\ddag}, Markus Aleksy\textsuperscript{$\diamond$}
}

\affil{\textsuperscript{*}Intelligent Networks, German Research Institute for Artificial Intelligence GmbH (DFKI), Trippstadter Straße 122, 67663 Kaiserslautern, Germany, Email: \{mathias.strufe, michael.gundall, hans.schotten\}@dfki.de}

\affil{\textsuperscript{\dag}Institut for Wireless Communication and Navigation, University of Kaiserslautern (TUK), Gottlieb-Daimler-Straße 47, 67663 Kaiserslautern, Germany, Email: schotten@eit.uni-kl.de}

\affil{\textsuperscript{\ddag}Nokia Bell Labs, Werinherstraße 91, \\ 81541 Munich, Germany, Email: \{christian.markwart, rakash.sivasivaganesan\}@nokia-bell-labs.com}

\affil{\textsuperscript{$\diamond$}ABB Corporate Research Germany, Ladenburg, Germany, Email: markus.aleksy@de.abb.com}

\abstractEn{The increasing demands for highly individual products as well as for flexible production lines represent new challenges. To address these demands, future plants must be highly flexible and dynamically reconfigurable. Current systems are usually based on wired technologies for the connection of sensors, actuators, and controlling or monitoring devices that allow only very limited dynamics. New applications, such as the use of robots, drones, or reconfigurable production lines, require the exploitation of wireless communication technologies.
However, current technologies are not able to meet the high requirements in terms of latency, robustness, resilience and data rate. The introduction of the 5th generation (5G) cellular communication system will meet these requirements for the first time.
Besides the use of radio-based solutions in new plants - so-called greenfield scenarios - deploying 5G also represents an efficient migration of existing plants - so-called brownfield scenarios - to Industry 4.0. 
In order to ensure that the challenging requirements are indeed meet in practical deployments of the new 5G technology, a tailor-made architecture is being developed within the Tactile Internet 4.0 (TACNET~4.0) project. As a basis for the design of the architecture, representative Industry 4.0 application scenarios, which are also be considered by the 3rd Generation Partnership Project (3GPP), were analyzed and compliance with the latest developments in the relevant standardization is also our target. 
The paper gives an overview of the considered use cases as well as the relevant reference architectures and the design process of the TACNET~4.0 architecture.}

\begin{document}

\maketitle

\section{Introduction}
Intelligent networks for fast and reliable data exchange are a key element of the raising Industry 4.0, since intelligent and connected machines need to exchange information directly with each other in real-time. In a Smart Factory, the production lines organize themselves independently and coordinate processes. This makes the production more flexible, dynamic and efficient \cite{[1]} \cite{[2]}. 

Wireless communications and in particular 5G will play a major role to achieve this paradigm change \cite{[3]} \cite{[4]}.
The three key features of 5G are: Enhanced Mobile Broadband (eMBB), ultra-reliable and low-latency communications (URLLC) and massive machine type communications (mMTC). This promise peak down- and uplink throughputs of 20 and 10~Gbps, while the end-to-end (E2E) latency is lower than 5~ms as well as up to one million connected devices per square kilometer.

This enables several new industrial use cases like remote controlled drones with high resolution video transmissions for industrial inspection, automated guided vehicles (AGVs) managed via a traffic control in the edge cloud to enable cooperative transport of large goods, thousands of new additional sensors for process automation, and predictive maintenance to name just a few examples. 

The seamless integration of these use cases in the already existing industrial communication infrastructure is one of the major goals of the TACNET~4.0 project \cite{[5]} that was initiated by the German Federal Ministry of Education and Research (BMBF) and combines 14 German industrial and academic partners. The goal of the project is the development of a unified industrial communication system with the continuous integration of 5G and other industrial communication networks. Existing industrial communication solutions will be integrated efficiently with the help of cross-network adaptation mechanisms and open interfaces between industrial and mobile communication systems.
The TACNET~4.0 architecture is therefore designed to be scalable in a technically and economically meaningful way and be usable for both large and small enterprises. The TACNET~4.0 consortium examined 20 industrial use cases \cite{[6]} in the area of Mobile Robotics, Local and Time Critical Control, Process Automation Monitoring, Remote Control, and Shared Infrastructure and analyzed them for the most critical communication requirements to design a unified industrial architecture, which integrates 5G wireless technologies and industrial communication networks to enable an efficient manufacturing. 

The remainder of this work is organized as follows. In section 2, a brief introduction of the already existing industrial reference architectures with their advantages and disadvantages is given. The four phases of the design process for the TACNET~4.0 architecture is described in section 3. Furthermore, in section 4 a SWOT analysis is done. Section 5 summarizes and concludes the work and gives an outlook on future work.

\section{Industrial Reference Architectures}

Various industrial reference architectures have taken part to establish a common Industrial Internet of Things (IIoT) architecture for the emerging Industry 4.0 use cases. The most relevant approaches are presented in this section.

\subsection{The Industrial Internet Reference Architecture (IIRA)}

First, the Industrial Internet Reference Architecture (IIRA) \cite{[7]}, which is build on top of the Industrial IoT Analysis Framework (IIAF), provides conventions, principles and practices for consistent description of IIoT architectures. IIRA documents the result of utilizing the IIAF to the IIoT systems. IIRA defines four viewpoints (business, usage, functional, and implementation) that address stakeholder-specific requirements. These viewpoints are complemented by two other dimensions: system life cycle process describing all process steps from conceptualization to disposal as well as the application scope, that is addressed industrial sectors. 

Moreover, the functional viewpoint is used to decompose an IIoT system into five functional domains to highlight its major building blocks, namely: control domain, operations domain, information domain, application domain, and business domain. It also describes the related functional components and the data and control flows in these domains as well as between them. The functional domains are complemented by two additional dimensions: system characteristics, such as safety, security, or reliability and crosscutting functions that are available across many of the system functional components, e.g. connectivity.

\subsection{Reference Architectural Model Industry 4.0 (RAMI 4.0)}
The Reference Architectural Model Industry 4.0 (RAMI 4.0) \cite{[8]} is an Industry 4.0-related model and consists of three dimensions: hierarchy levels, life cycle \& value stream, and layers. Hierarchy levels cover the required functionalities by a factory or entire plant. They are based on the IEC 62264 \cite{[9]} / IEC 61512 \cite{[10]} standards and extend them by elements “product” and “connected world”. Life cycle \& value stream is the second dimension used in the model. It considers IEC 62890 \cite{[11]} and reflects the life cycle of products and machines supporting types as well as instances. The layers describe the IT-based elements of the system in a structured way. They start with a business perspective and end on asset level.

\subsection{one M2M Service Platforms}
The Functional Architecture for the oneM2M Services Platform (oneM2M) from European Telecommunications Standards Institute (ETSI) is specified in \cite{[12]} and defines a 3-layer model to support E2E machine-to-machine (M2M) services. oneM2M provides a common framework for interoperability between the many M2M and IoT technologies being introduced. Especially with the second and third release, oneM2M opens the IoT ecosystem to devices that lack the protocol and enables also the interworking among systems based on alternative approaches like AllSeen Alliance’s AllJoyn, Open Connectivity Foundation’s OIC, and the Open Mobile Alliance’s Lightweight M2M (LWM2M). 

\subsection{Reference Architecture Conclusion}
While the proposed industrial architectures have a good focus on the general design of industrial internet systems, interoperability between multiple IoT technologies and covering the full life cycle of manufacturing, none of them focuses on wireless communication. 
This leads to the decision to develop an own tailored architecture.

\section{Architecture Design Process}

\begin{figure*}[!ht]
  \includegraphics[width=6.7in,height=1in]{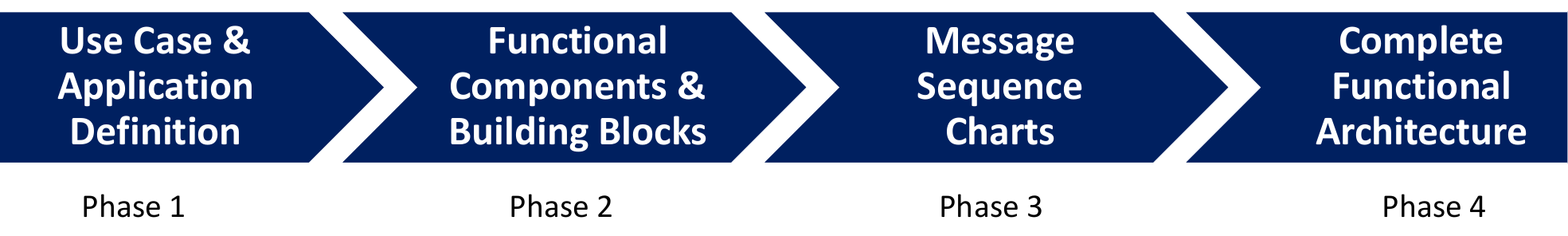}
  \caption{Architecture Development Process.}
  \label{ArchDevProcess}
\end{figure*}

The architecture development process comprises four design phases, shown in figure \ref{ArchDevProcess}. 

The first step is to define the targeted applications and their requirements that the architecture needs to handle. The second phase compromises the definition of functional components including needed in- and output metrics to realize the use cases. With the help of the message sequence charts, the interactions between single functions get described and also missing interfaces get identified. After an iterative process of phase 2 and 3, the result is a complete functional architecture.

\subsection{TACNET4.0 Industrial Use Case Definition}
The first phase comprises the definition of the use cases that are considered within the TACNET~4.0 project and their requirements in particular on a wireless communication system. 

\begin{wrapfigure}{r}{9cm}
  \includegraphics[width=3.5in,height=2.1in]{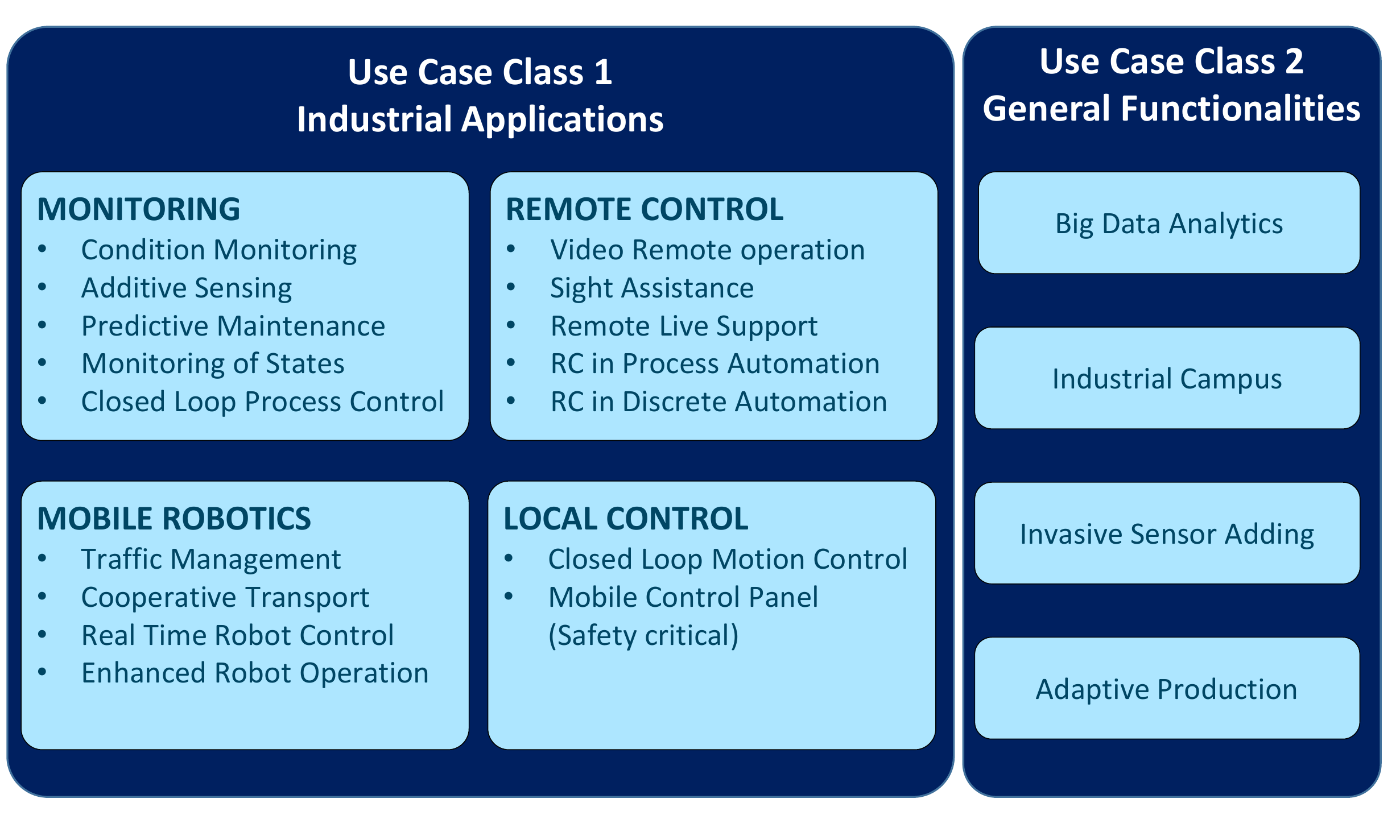}
  \caption{TACNET~4.0 Use Case Classes.}
  \label{UCC}
\end{wrapfigure}

In this phase, especially the industrial partner of the TACNET~4.0 consortium examined 20 most relevant industrial use cases which can be divided into two use case classes: Industrial Application and General Functionalities as shown in figure \ref{UCC}. 

The use case class 1 industrial use cases again can be mapped into four main functional groups: Monitoring, Remote Control, Local Control and Mobile Robotics shown in figure \ref{UCC}.

For example, Additive Sensing is from high interest in the field of monitoring. The idea is to add with low expenditures wireless communicating and battery powered sensors to all critical points in your infrastructure to collect data and make predictive maintenance possible. This can raise the number of connected sensors into the thousands per production hall.

Remote Live support is one of the most prominent use cases in Remote Control. The maintainer on site can get the help of a remote expert and get direct helpful information in his view via Augmented Reality (AR) glasses. Since the calculation of view in the high resolution video is done in the edge cloud high bandwidth and very low latency is needed to avoid cybersickness.

In terms of Local Control operating machines via a mobile control panel provides higher flexibility, however, the critical element of such a control panel is the availability of an emergency stop, which has to operate to the according to the strict safety standards also over a wireless link. Mobile Robotics will also play a major role in the factory of the future, these need a Traffic Management in the edge cloud to achieve cooperative transport of large goods or platooning.

The use case class 2 General Functionality includes commonly needed functions that needed when running one or multiple industrial applications. Typical use cases in this class are Big Data Analytics and visualization of Quality of Service (QoS) Data, sensor logs and security relevant data for predictive maintenance and anomaly detection. Also considered here are network functions for shared private network infrastructure so-called Industrial Campus.

\subsection{Functional Components}

In the second phase, the needed functional components get determined to fulfill the requirements of the defined use cases as well as common preliminary functions during the bootstrapping of the system.

For example of a preliminary function is the Initial Registration Process to connect completely new devices the first time to the TACNET~4.0 system.
In the area of security there are  Authentication and Authorization functions needed. Furthermore Time Synchronization function is important to sync all network clocks to provide deterministic communication. A Localization service is essential for the Remote Live Support or Traffic Management of the AGVs. Spectrum Monitoring and Management Function are important to regulate the interaction of the different radio access technologies to just name a few.  

In this phase also the required input and provided output of each function get described in a high level way.

\subsection{Message Sequence Charts}

In a third step, the developed functions are described using Message Sequence Charts to identify the required interfaces.

\begin{figure*}[!ht]
  \includegraphics[clip, trim=0.0cm 8.0cm 0.0cm 0.0cm, width=\textwidth]{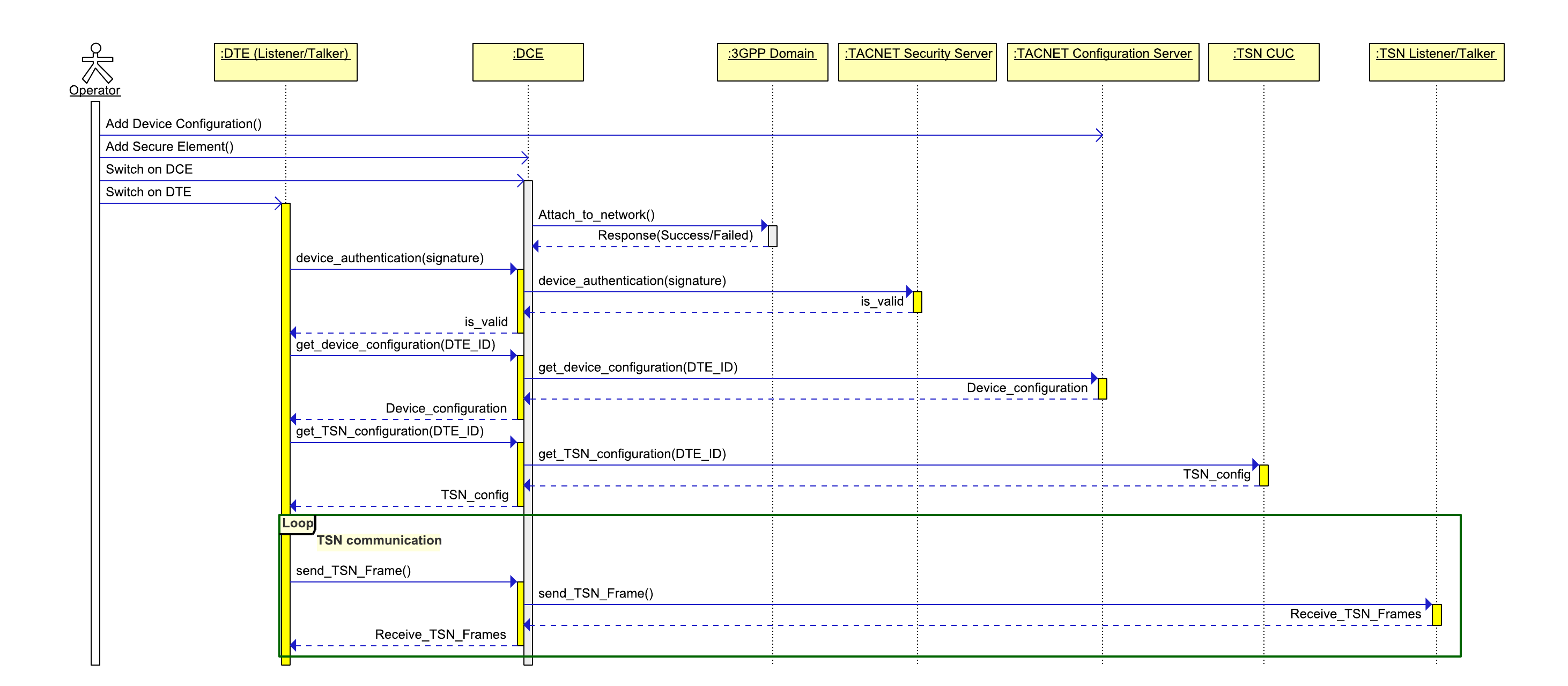}
  \caption{Initial registration Process.}
  \label{init}
\end{figure*}

As one example: for the Initial Registration, the Network Operator first needs to manual add the configuration of the new device (e.g. sensors / actuators) in the configuration server and also add the secure element (e.g. SIM Card) to the data communication equipment (DCE) and data terminal equipment (DTE), compare figure \ref{init}. 

After that, the DTE, which serves as the interface to the physical sensor/actuator or bus system (e.g. Profinet), and the DCE can be switched on. The DCE then first establishs the connection via the air interface to the 3GPP core authentication service. After the successful radio connection is established, the signature of the DTE is checked by the TACNET~4.0 internal authentication and authorization function, to check, if access is granted and to which systems the new device has access. Only then, the DTE can retrieve its initial settings from the configuration server. In case that the DTE is a TSN End Device, it now needs also to register at the Centralized User Configuration (CUC) that than defines the type of the TSN transmission (e.g. E2E communication). After that, the DTE is finally ready to transmit TSN frames.

\subsection{Complete Functional Architecture}

After several iteration and refinements in phase 2 and 3, phase 4 bring all single components together.
To provide maximum scalability, the TACNET~4.0 architecture is divided into five layers: User Plane, Control Plane, Management \& Orchestration Plane, Application \& Service Plane, as well as a Security Plane.

\begin{figure*}[!ht]
  \includegraphics[width=\textwidth]{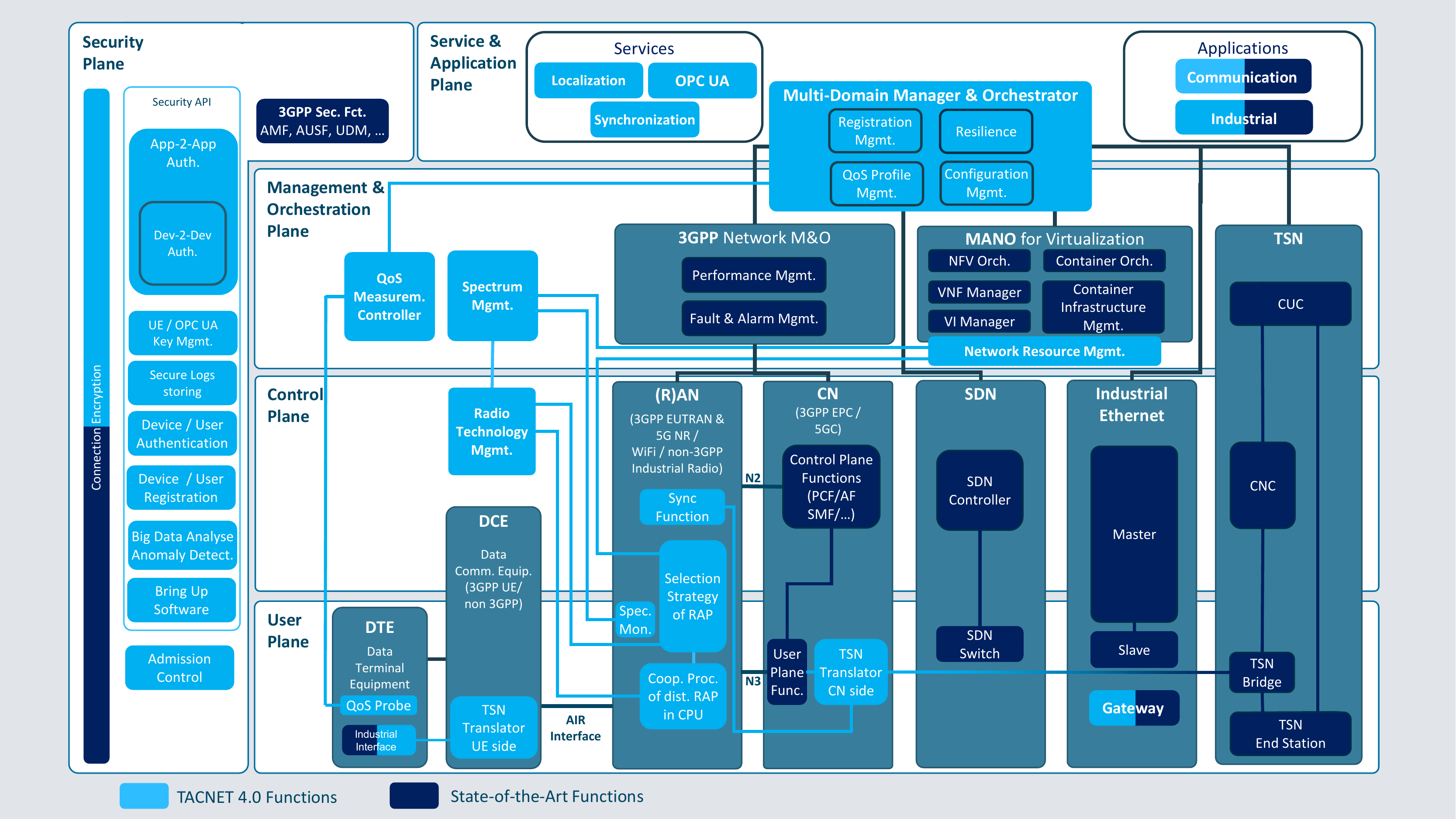}\label{arch}
  \caption{TACNET 4.0 Architecture.}
\end{figure*}

The User and Control plane comprise the 5G radio access network (RAN), the 3GPP Core Network (CN) \cite{[13]}, the Software-Defined Network (SDN) as well as the Industrial Ethernet (IE). The Time-Sensitive Network (TSN) \cite{[14]} is also located in these planes, but also extend to the Management and Orchestration Plane, since the CUC of the TSN needs to manage the TSN end devices and make requests to the Centralized Network Controller (CNC) to establish deterministic communication with specific requirements. 

Also in the Management and Orchestration Plane located are the Management and Network Orchestration (MANO) for the virtualized network functions (VNF) and container management, as well as the 3GPP Network Management and Orchestration which is in charge of the Performance, Fault and Alarm Management of the 3GPP CN and RAN.

Since one of the main targets of TACNET~4.0 is the seamless integration of existing industrial communication solutions and 5G in one architecture a higher level management instance is needed, the so-called Multi-Domain Manager \& orchestrator. This component is overall responsible for the registration of new devices and the configuration of devices and network technologies to achieve a constant QoS across multiple domains.

Furthermore, the identified Communication and Industrial Applications as well as needed Services (e.g. synchronization or localization) to fulfill the already defined use cases are located on the top, in the Service and Application Plane. 

Since security is of great importance, an additional Security Plane with functions for Device and User Authentication, secure Log storing comprising all other four layers and also provide encrypted connections for a secure communication within the system.

\section{SWOT analysis}

The Strengths, Weaknesses, Opportunities, and Threats (SWOT) analysis helps to identify the current position of the project and develop a future business strategy.

\begin{table}[!ht]
  %\begin{tabularx}{\hsize}{|p{0.2cm}|X|X|}\hline
  \begin{tabularx}{\hsize}{|X|X|}\hline
    Strengths 
    & 
    Weaknesses 
    \\\hline
    
    %\begin{turn}{90}Opportunities\end{turn}
    \begin{itemize} 
        \item Design of an holistic architecture, 
        \item Considering cutting edge technologies (5G, TSN, SDN),
        \item Strong consortium consisting of a good mix of industrial and academic partners, 
        \item Scalable and extensible architecture; 
    \end{itemize} 
    &  
    \begin{itemize} 
        \item Technology risk, new technologies at an early development stage can affect deadlines,
        \item Architecture may get to complex for small size enterprises;
    \end{itemize} 
    \\\hline

    Opportunities
    & 
    Threats
    \\\hline

    \begin{itemize} 
        \item First holistic architecture for industrial environments considering 5G,
        \item Strengthening German small- and medium size-enterprises (SMEs);
    \end{itemize}       
    &   
    \begin{itemize} 
        \item Various competitive architectures;

    \end{itemize}  \\\hline

  \end{tabularx}
  \caption{The SWOT analysis.}\label{SWOT}
\end{table}

Table \ref{SWOT} shows that the strength of TACNET~4.0 is the well-chosen consortium with a good mixture of universities, research centers, equipment vendors, and specialized SMEs. That allows establishing links to relevant
stakeholders communities. Furthermore, the wide range of experience and balance between academic and industrial partners within this consortium help to cover a broad spectrum of well known and cutting edge technologies, which results in a holistic, scalable and extensible architecture.
This opens the opportunity to be the first holistic architecture considering 5G on the market and boost German industry sector.

But there are also some weaknesses and risks. Considering new cutting edge technologies can also delay the whole project, if an elementary technology gets delayed. To avoid this, each TACNET~4.0 partner is continuously monitoring the status of new technologies in his field of expertise and designs the architecture as modular as possible to replace discontinued technologies if necessary. A modular architecture design is also the solution to adjust the architecture to the needs and reduce thereby the complexity and make it suitable even for small- and medium-size enterprises. 
The external risk, the presence of competitive industrial architectures, can also be mitigated by the strong consortium with global players and good insight into market developments, as well as presence in several standardization bodies.

\section{Summary and Outlook}

This paper presents the design process of the TACNET~4.0 architecture. Starting from a brief overview of competing industrial architectures, it describes the whole design process from the use case identification and communication requirement specification, via derivation of functional components, up to the development of Message Sequence Charts to the full functional architecture.
In the next step the developed architecture will be experimental evaluated in a testbed.

\section{Acknowledgement}
This work has been supported by the Federal Ministry of Education and Research of the Federal Republic of Germany as part of the TACNET~4.0 project with grant number 16KIS0715.

\end{document}